\documentclass[11pt]{article}
\usepackage{epsfig}
\usepackage{amsmath,amssymb,amsfonts,phonetic}
\usepackage{psfrag}
\usepackage{enumitem}
\usepackage{color}
\usepackage[colorlinks]{hyperref}
\usepackage{caption}
\usepackage{epstopdf}

\begin{document}

\title{\textbf{Fragmentation of Fast Josephson Vortices and Breakdown of Ordered States by Moving Topological Defects} }

\author{Ahmad Sheikhzada$^*$ and Alex Gurevich$^{**}$ \let\thefootnote\relax\footnote{Department of Physics and the Center 
for Accelerator Science, Old Dominion University, Norfolk, VA 23529, USA. $^*$e-mail: ashei003@odu.edu,  $^{**}$e-mail: gurevich@odu.edu}}

\date{\vspace{-5ex}}

\maketitle

\begin{abstract}
\textbf{Topological defects such as vortices, dislocations or domain walls define many important effects in superconductivity, superfluidity, magnetism, liquid crystals, and  plasticity of solids. Here we address the breakdown of the topologically-protected stability of such defects driven by strong external forces. We focus on Josephson vortices that appear at planar weak links of suppressed superconductivity which have attracted much attention for electronic applications, new sources of THz radiation, and low-dissipative computing.  Our numerical simulations show that a rapidly moving vortex driven by a constant current becomes unstable with respect to generation of vortex-antivortex pairs caused by Cherenkov radiation. As a result, vortices and antivortices become spatially separated and accumulate continuously on the opposite sides of an expanding dissipative domain. This effect is most pronounced in thin film edge Josephson junctions at low temperatures where a single vortex can switch the whole junction into a resistive state at currents well below the Josephson critical current.  Our work gives a new insight into instability of a moving topological defect which destroys global long-range order in a way that is remarkably similar to the crack propagation in solids. }
\end{abstract}

Quantized vortex lines are quintessential topological defects \cite{mermin,topol} which determine the behavior of superconductors and superfluids. Vortices in superconductors are characterized by an integer winding number $n$ in the phase $\varphi$ of the complex order parameter $\Psi=\Delta\exp(i\varphi)$, singularity of $\nabla\varphi$ in a vortex core, and the quantized magnetic flux, $\phi=n\phi_0$, where  $\phi_0= h/2e=2.07\times 10^{-15}$ Wb is the magnetic flux quantum, $e$ is the electron charge and $h$ is the Planck constant. Because destruction of a topological defect requires overcoming a huge energy barrier proportional to the length or the area of a system, vortices can only disappear by annihilating with antivortices with the opposite sign of $n$ or exiting through the sample surface, or forming shrinking loops. This brings about the question: does this fundamental, topologically-protected stability of vortices remain preserved for a vortex driven by a strong force and, more generally, what happens to a global long-range order if a moving topological defect becomes unstable? To address this issue, we performed numerical simulations of vortices in superconducting-insulating-superconducting (SIS) Josephson junctions \cite{BP,KKL} where the lack of suppression of the superconducting gap $\Delta(\textbf{r})$ greatly reduces viscous drag of vortices, allowing them to reach velocities as high as a few percent of the speed of light $c$ under a strong current drive.  Dynamics of superfast Josephson vortices has been probed in annular thin film junctions \cite{ustinov}, and  has recently attracted much attention for the development of superconducting qubits and low-dissipative digital memory \cite{comp,qc,ust}, and new sources of THz radiation \cite{thz}.  We show that a rapidly moving vortex can become unstable, causing a cascade of  expanding vortex-antivortex pairs, which eventually destroy the global long-range order.  This effect may impose limitations on the performance of Josephson memory \cite{comp,qc,ust}, superconducting sources of THz radiation \cite{thz}, or polycrystalline superconducting resonator cavities for particle accelerators \cite{acc}, and have broader  implications for other systems with long-range order.

We start with a standard theory of a Josephson vortex in a long junction described by the sine-Gordon equation for the phase difference of the order parameter $\theta(x,t)=\varphi_1-\varphi_2$ between two bulk electrodes \cite{BP,KKL}:  
\begin{equation}
\ddot{\theta}+\eta\dot{\theta} = \theta'' -\sin \theta+\beta.
\label{sg}
\end{equation}
Here the prime and the overdot denote partial derivatives with respect to the dimensionless coordinate $x/\lambda_J$ and time $\omega_J t$,  
$\omega_J=\left(2\pi J_c/\phi_0 C\right)^{1/2}$ is the Josephson plasma frequency, $J_c$ is the tunneling critical current density, $C$ is the specific capacitance of the junction, $\lambda_J=\left(\phi_0/4\pi\mu_0\lambda J_c\right)^{1/2}$ is the Josephson penetration depth, $\lambda$ is the London penetration depth, $\eta = 1/\omega_J RC$ is the damping constant due to the ohmic quasiparticle resistance $R$, and $\beta=J/J_c$ is the driving parameter controlled by a uniform transport current density $J$.  

The sine-Gordon equation has been one of the most widely used equations to describe topological defects in charge and spin density waves \cite{cdw}, commensurate-incommensurate transitions \cite{com1,com2,com3}, magnetic domain walls \cite{mag}, dislocations in crystals \cite{disl,sg}, kinks on DNA molecules \cite{dna1,dna2}, etc. Particularly, the $2\pi$ kink solution $\theta(x,t) = 4\tan^{-1}\exp[(x-vt)/L]$ of equation (\ref{sg}) at $\eta\to 0$ describes a Josephson vortex of width $L=(1-v^2/c_s^2)^{1/2}\lambda_J$ moving with a constant velocity $v$, where $c_s=\omega_J\lambda_J$ is the Swihart velocity of propagation of electromagnetic waves along the junction \cite{BP}. As $v$ increases, the vortex shrinks at $\eta\ll 1$ and expands at $\eta>1$ \cite{KKL}. 

The Lorentz-like contraction of the Josephson vortex  at $\eta\ll1$ indicates that equation (\ref{sg}) should be modified at large velocities if $L(v)$ approaches the geometry-dependent magnetic screening length $\Lambda$. Indeed, equation (\ref{sg}) was obtained assuming that both $\theta(x,t)$ and the magnetic field $B(x,t)$ produced by vortex currents vary slowly along the junction over the same length $\sim L(v)\gg \Lambda$ \cite{BP}; otherwise $\theta(x,t)$ and $B(x,t)$ vary over {\it different} lengths and the relation between $B(x,t)$ and $\theta(x,t)$ becomes nonlocal \cite{KKL}. The equation, which generalizes  equation (\ref{sg}) to $\theta(x,t)$  and $B(x,t)$ varying over any lengths larger than the superconducting coherence length $\xi$, is given by  \cite{ivan,AG,rgm,abdu}:         
\begin{gather}
\ddot{\theta}+\eta\dot{\theta} = \epsilon\int_{-\infty}^{\infty} G\left(\frac{|x-u|}{\alpha}\right)\frac{\partial ^2\theta}{\partial u^2} du -\sin\theta+\beta,
\label{int}
\end{gather} 
where $\epsilon=\lambda_J/\lambda$, $\alpha = \Lambda/\lambda_J$, and the kernel $G(x)$ depends on the sample geometry. Here  $G(x/\alpha)=\pi^{-1}K_0(x/\alpha)$ for a planar junction in a bulk superconductor, where  $\alpha=\lambda/\lambda_J$ and $K_0(x)$ is the modified Bessel function \cite{AG}. For an edge junction in a thin film of thickness $t\ll\lambda$, we have $2G(x/\alpha)=\textbf{H}_0(x/\alpha)-Y_0(x/\alpha)$, where $\alpha=2\lambda^2/t\lambda_J$,  and $\textbf{H}_0(x)$ and $Y_0(x)$ are the Struve and Bessel functions, respectively \cite{rgm,abdu}. The kernels $G(x,u)$ for different geometries decrease with $u$ at $|x-u|>\Lambda$ and have the same logarithmic singularity at $u=x$ \cite{ivan,AG,rgm,abdu}.  The nonlocal effects are most pronounced at $\lambda_J^2/\lambda < \Lambda$, particularly in high-$J_c$ bulk junctions with $J_c > J_d/\kappa$ \cite{AG} and thin film junctions \cite{ivan,rgm,abdu,krasnov}  with large Pearl length $\Lambda=2\lambda^2/t$, where $J_d = \phi_0/2^{3/2}\mu_0\lambda^2\xi$ is the depairing current density, and $\kappa=\lambda/\xi$. At $\lambda^2_J/\lambda\ll\Lambda$, only the universal logarithmic part of $G(x,u)$
\begin{equation}
G_0(|x-u|/\alpha)=\pi^{-1}\ln(\alpha/|x-u|)
\label{log}
\end{equation}
is essential, while a nonsingular, geometry-dependent part of $G(x,u)$ can be disregarded \cite{AG,rgm,abdu}. Equations (\ref{int})-(\ref{log}) describe mixed Abrikosov vortices with Josephson cores of length $l=\lambda^2_J/\lambda\simeq \xi J_d/J_c$ along the junction (AJ vortices) \cite{AG}. Equations (\ref{int})-(\ref{log}) in the overdamped limit of $\eta\gg 1$ have an exact solution $\theta(x,t)=\pi+\sin^{-1}\beta+2\tan^{-1}[(x-vt)/L]$ that describes a driven AJ vortex core with weak suppression of $\Delta(x)$  and the length $L=(1-\beta^2)^{-1/2}l$ expanding as $\beta$ increases \cite{AG}.  AJ vortices have been observed in flux flow experiments on low-angle grain boundaries of high-$T_c$ cuprates \cite{gb}, the $c-$axes resistivity in layered pnictides  \cite{moll}, and most recently by STM imaging of step edge junctions in Pb and In monolayers on Si substrates \cite{pb,in}. Equations (\ref{int})-(\ref{log}) also reduce to the Peierls equation that describes slow dislocations in crystals \cite{disl}.
  
Unlike the sine-Gordon equation, the nonlocal equation (\ref{int}) at $\eta=0$ is not Lorentz-invariant, so a uniformly moving vortex can radiate Cherenkov waves $\delta\theta(x,t) \propto \exp(ikx-i\omega_k t)$ with the phase velocities $\omega_k/k$ smaller than $v$ \cite{abdu,MS}. The condition of Cherenkov radiation at $\eta=0$ is given by:
\begin{equation}
kv > \omega_J\left[ \sqrt{1-\beta^2}+lk^2G(k) \right]^{1/2},
\label{cher}
\end{equation}
where $l=\lambda_J^2/\lambda=\phi_0/4\pi\mu_0\lambda^2 J_c$, and $G(k)$ is the Fourier image of $G(x)$.  Here $G(k)$ decreases as $1/k$ at $k>\Lambda^{-1}$ so equation (\ref{cher}) is satisfied for $k>k_c$ where the maximum wavelength $2\pi/k_c$ increases with $v$  \cite{suppl}.   To address the effect of Cherenkov radiation on the moving vortex, we performed numerical simulations of equation (\ref{int}) for SIS junctions of different geometries.  

Shown in Figure \ref{panel1} are the numerical results for a planar bulk junction at $\eta=0.05$ and the large ratio $\lambda_J/\lambda = 10$ usually described by the sine-Gordon equation (\ref{sg}). Yet the more general integral equation (\ref{int}) reveals the effects which are not captured by equation (\ref{sg}), particularly a trailing tail of Cherenkov radiation behind a vortex moving with a constant velocity \cite{MS}. Moreover, as the amplitude and the wavelength of radiation increase with $v$, the vortex becomes unstable at $\beta>\beta_s$, the instability is triggered at the highest maximum of Cherenkov wave where $\theta_m$ reaches a critical value $\theta_c\approx 8.65-8.84$, depending on $\eta$, $\lambda/\Lambda$, and the junction geometry \cite{suppl}. Here $\theta_c$ is confined within the interval $5\pi/2 <\theta_c < 3\pi $ in which a uniform state of a Josephson junction is unstable \cite{BP,KKL}. As the velocity increases, the domain where  $5\pi/2 <\theta(x-vt) < 3\pi $ behind the moving vortex widens and eventually becomes unstable as its length exceeds a critical value.  This suggests a qualitative picture of the vortex instability caused by the appearance of a trailing critical nucleus being in the unstable $\pi$-junction state \cite{BP, KKL} caused by strong Cherenkov radiation. The latter appears entirely due to the Josephson nonlocality described by  equation (\ref{int}), which  has no steady-state vortex solutions at $J>J_s$ where $J_s$ can be well below $J_c$ at which the whole junction switches into a resistive state. 

The dynamic solutions of equation (\ref{int}) at $\beta>\beta_s$ change strikingly.  Our simulations have shown that the instability originates at the highest maximum $\theta=\theta_m$ of the trailing Cherenkov wave which starts growing and eventually turning into an expanding vortex-antivortex pair \cite{suppl}, as shown in Figure \ref{panel2}. As the size of this pair grows, it generates enough Cherenkov radiation to produce two more vortex-antivortex pairs which in turn produce new pairs. Continuous generation of vortex-antivortex pairs results in an expanding dissipative domain in which vortices accumulate at the left side, antivortices accumulate at the right side, while dissociated vortices and antivortices pass through each other in the middle \cite{suppl}. As a result, $\theta(x,t)$ evolves into a growing ``phase pile'' with the maximum $\theta_m(t)$ increasing approximately linear with time and the edges propagating with a speed which can be both smaller and larger than $c_s$, the phase difference $\theta(\infty)-\theta(-\infty)=2\pi$ between the edges remains fixed.  We observed the phase pile dynamic state for different junction geometries and $\eta$ ranging from $10^{-3}$ to $0.5$ \cite{suppl}. For instance,  Figures \ref{3d} and \ref{te3d} show the 3D images of the initial stage of dynamic separation of vortices and antivortices  calculated for a bulk junction and a thin-film edge junction. Here the local magnetic field $B(x,t)$ oscillates strongly at the moving domain edges but becomes rather smooth away from them, as shown in Figure \ref{field}. In the most part of the phase pile overlapping vortices are indistinguishable, yet  the net flux $\phi=\phi_0$ of this evolving multiquanta magnetic dipole remains quantized.  

Shown in Figure \ref{iv} are the steady-state vortex velocities $v(\beta)$ calculated for different junction geometries. The instability corresponds to the endpoints of the $v(\beta)$ curves which  have two distinct parts. At small $\beta \lesssim \eta$ the velocity $v(\beta)$ increases sharply with a slope limited by a weak quasiparticle viscous drag. At larger $\beta\gtrsim\eta$ the increase of $v(\beta)$ with $\beta$ slows down, as the vortex velocities are mostly limited by radiation friction \cite{MS} and depend weakly on the form of dissipative terms in equation (\ref{int}).  For a low-$J_c$ junction with $\lambda_J/\lambda=10$, the effect of Cherenkov radiation on $v(\beta)$ is weak, but for a high-$J_c$ bulk junction with $\lambda/\lambda_J = \sqrt{10}$ and $\eta\ll 1$, radiation friction dominates at practically all $\beta$, significantly reducing both $v(\beta)$ and $\beta_s$.

For thin film edge junctions, the critical splitting current density $J_s$ gets reduced down to $J_s\approx 0.4J_c$ at $\eta = 10^{-3}$, as shown in Figure \ref{iv}.  In the extreme nonlocal limit described by equations (\ref{int}) and (\ref{log}), the maximum velocity $v_s=v(J_s)$ at which the steady-state moving vortex remains stable, can be written in the scaling form $v_s=c_s\lambda_Jf(\eta)/\lambda$, where $f(\eta)$ decreases from $\simeq 2.5$ at $\eta=0.5$ to $\simeq 1$ at $\eta=10^{-3}$. The Josephson vortex in thin film edge junctions can reach the velocities exceeding the nominal Swihart velocity $c_s=\omega_J\lambda_J$ at $J\simeq J_s$ if $\lambda_J > \lambda$ but $l<\Lambda$, that is, $t<2\lambda^3/\lambda_J^2$. Dynamics of $\theta(x,t)$ in the nonlocal limit at $J>J_s$ is similar to that is shown in Figures 1-3, except that the edges of phase pile can propagate with  ``superluminal''  velocities $v\simeq v_s>c_s$ if $\lambda_J>\lambda$ \cite{suppl}.  Once vortex-antivortex pairs start replicating, the speed of leading vortices at the edges gradually increases from $v_s$ to a limiting value $v_\infty$, for instance, from $v_s\approx 0.72 l\omega_J$ to $v_\infty\approx 1.12 l\omega_J $ for an edge junction with $l=\Lambda/2$ and $\eta=0.1$ \cite{suppl}.   

The effects reported here are most pronounced in underdamped SIS junctions between s-wave superconductors at low temperatures for which the viscous drag coefficient $\eta\propto \exp(-\Delta/T)$ due to thermally-activated quasiparticles \cite{BP} is small. Here $\eta\ll 1$ also implies that a moving vortex does not generate additional quasiparticles because the induced Josephson voltage $V= v\hbar\theta'_m/2eL$ is smaller than $\Delta/e$, where $\theta_m'$ is the maximum phase gradient. These conditions are satisfied for the parameters used in the calculations and are facilitated by the electromagnetic nonlocality of thin film edge junctions \cite{suppl}, particularly monolayer junctions \cite{pb,in}. Furthermore, a small power $P$ dissipated by a moving vortex at $\eta\ll 1$ does not really affect the Cherenkov instability. For instance, $P$ generated by a vortex at the critical velocity $v_s\simeq \omega_Jl$ in a thin film junction is given by \cite{suppl}:       
\begin{equation}
P\simeq \frac{\phi_0^2 t}{4\pi\mu_0\lambda^2 RC}
\label{P}
\end{equation}
This equation shows that the power $P$ is independent of $J_c$ and is greatly reduced in the underdamped limit at low temperatures as the quasiparticle resistance $R$ of SIS junctions becomes exponentially large at $T\ll T_c$. To estimate $P$, it is convenient to write equation (\ref{P}) in the form $P\simeq \eta\varepsilon_0 t\omega_J$, where $\varepsilon_0= \phi_0^2/4\pi\mu_0\lambda^2$ is a characteristic line energy of Abrikosov vortex \cite{hts}. For an edge junction in a Nb film with $t=1$ nm, $\lambda=40$ nm, $\varepsilon_0\sim 10^4$ kelvin/nm, and 
$\omega_J = 100$ GHz much smaller than $\Delta/\hbar\simeq 2.4$ THz \cite{acc},  equation (\ref{P}) yields $P\sim 0.16$ nW at $\eta=10^{-2}$. Local overheating $\delta T=PY_K$ caused by vortex dissipation is further reduced in thin film junctions for which the energy transfer to the substrate due to ballistic phonons is much more effective than diffusive phonon heat transport in thick samples, where $Y_K$ is the Kapitza interface thermal resistance \cite{gm}. Such weak overheating caused by a moving vortex cannot result in thermal bistability and hysteric switching due to hotspot formation \cite{gm}.

Proliferation of vortex-antivortex pairs triggered by a moving Josephson vortex can be essential for the physics and applications of weak link superconducting structures where the formation of expanding phase pile patterns can switch the entire junction into a normal state at currents well below the Josephson critical current, $J>J_s\simeq (0.4-0.7)J_c$. Such dynamic vortex instability can result in hysteretic jumps on the V-I curves which appear similar to those produced by  heating effects \cite{KKL,thz}, yet this instability is affected by neither cooling conditions nor the nonequilibrium kinetics of quasiparticles.  Indeed, heating is most pronounced in overdamped junctions with $\eta > 1$  in which Cherenkov radiation is suppressed. By contrast, the Cherenkov instability is characteristic of the weakly-dissipative underdamped limit $\eta\ll 1$, although Figure \ref{iv} shows that this instability in thin film edge junctions can persist up to $\eta = 0.5$. Therefore, the crucial initial stage of the phase pile formation at $\eta\ll 1$ is unaffected by heating which may become more essential at the final stages of the transition of the entire junction into the normal state. At $\eta\sim 1$ the Cherenkov instability may be masked by heating effects, particularly in bulk junctions for which heat transfer to the coolant is less efficient than in thin films. 

It should be emphasized that the instability reported here does not require special junctions with $J_c\sim J_d$. In fact, even for the seemingly conventional bulk junction with $\lambda_J = 10 \lambda$ shown on the top panel of Figure \ref{iv}, the instability at $J_s\simeq 0.846J_c$ implies $J_c \sim 0.01J_d/\kappa$, which translates into $J_c \sim 10^{-4}J_d$ for bulk NbN junctions. Moreover, in wide thin film edge junctions the nonlocality becomes important at even much lower $J_c$, as is evident from the lower panel of Figure \ref{iv}. Therefore, the effects reported here can occur in conventional underdamped junctions with $J_c\ll J_d$, particularly wide thin film or monolayer edge junctions. Interaction of Josephson vortices with pinned Abrikosov vortices in electrodes can result in additional mechanisms of splitting instability of Josephson vortices.  For instance, radiation by Josephson vortices can be greatly enhanced as they move in a periodic magnetic potential of Abrikosov vortices along weak link grain boundaries \cite{gb,gc}, whereas Abrikosov vortices trapped perpendicular to the Josephson junction can result in generation of Josephson vortex-antivortex pairs in the presence of the applied electric current \cite{mil}. 

Our results can be essential for other topological defects such as crystal dislocations or magnetic domain walls described by the generic nonlocal equation (\ref{int}) in which the integral term results from a common procedure of reduction of coupled evolution equations for several relevant fields to a single equation.  For Josephson junctions, such coupled fields are $\theta$ and $B$, but for domain walls in ferromagnets, the nonlocality can result from long-range magnetic dipolar interactions \cite{magdw}. For dislocations, the nonlocality and Cherenkov radiation of sound waves in equation (\ref{int}) come from the discreteness of the crystal lattice \cite{sg} and long-range strain fields \cite{disl},  although the dynamic terms in the Peierls equation \cite{dis1,dis2} are more complex than those in equation (\ref{int}).   Dynamic instabilities of dislocations have been observed in the lattice Frenkel-Kontorova models \cite{sg} in which sonic radiation can also result from periodic acceleration and deceleration of a dislocation moving in a crystal Peierls-Nabarro potential \cite{disl}. The latter effect becomes more pronounced as the dislocation core shrinks at higher velocities and becomes pinned more effectively by the lattice. By contrast, the instability reported here results entirely from Cherenkov radiation, the condition (\ref{cher}) can be satisfied for any system in which $G(k)$ in equation (\ref{cher}) decreases with $k$. This instability can thus have broader  implications: for instance,  the phase pile dynamics of Josephson vortices appears similar to a microcrack propagation caused by a continuous pileup of subsonic dislocations with antiparallel Burgers vectors at the opposite tips of a growing crack described by equations (\ref{int}) and (\ref{log}) \cite{disl}. 

Our results give a new insight into breakdown of a global long-range order which has been usually associated with either thermally-activated proliferation of topological defects (like in the Berezinskii-Kosterletz-Thouless transition) or static arrays of quenched topological defects pinned by the materials disorder \cite{topol}. Here we point out a different mechanism in which a long-range order is  destroyed as a single topological defect driven by a strong external force becomes unstable and triggers a cascade of expanding pairs of topological defects of opposite polarity.

\pagebreak

\section*{Methods}
We have developed an efficient MATLAB numerical code to solve the main integro-differential equation (\ref{int}) using the method of lines \cite{lines}. By discretizing the integral term in equation (\ref{int}) it was reduced to a set of coupled nonlinear ordinary differential equations in time which were solved by the multistep, variable order Adams-Bashforth-Moulton method \cite{ode}. We have  checked our numerical results using a slower iterative method to make sure that the logarithmic singularity of $G(x-u)$ is handled properly, the absolute and relative error tolerances were kept below $10^{-6}$.  The length $L_b$ of computational box $x_1<x<x_1+L_b$ along the $x-$axis (either co-moving with the vortex or expanding  with the phase pile) was taken large enough to assure no artifacts coming from possible reflected waves at $x=x_1$ and $x=x_1+L_b$.  We set $\theta(x_1,t) - \sin^{-1}\beta < 10^{-6}$ and $\theta(x_1+L_b,t)-2\pi-\sin^{-1}\beta < 10^{-6}$ and made sure that changing $L_b$ does not affect the results, where $L_b$ was typically taken at least three times larger than the spatial extent of $\theta(x,t)$, be it a single vortex or expanding phase pile. The steady state phase distribution $\theta(x-vt)$ in a uniformly moving vortex at a given $\beta$ was computed by solving the full dynamic equation (\ref{int}) using the single-vortex solution calculated at a smaller preceding value of $\beta$ as an  initial condition. The code then run until the velocity of the vortex stabilizes to the accuracy better than 0.1$\%$.

\pagebreak

\section*{Acknowledgments}
This work was supported by the United States Department of Energy under Grant No. DE-SC0010081.

\section*{Supplementary Information}
For supplementary information and simulations see http://www.nature.com/articles/srep17821.

\break

\begin{figure}[t]
\centering\includegraphics[width=\textwidth]{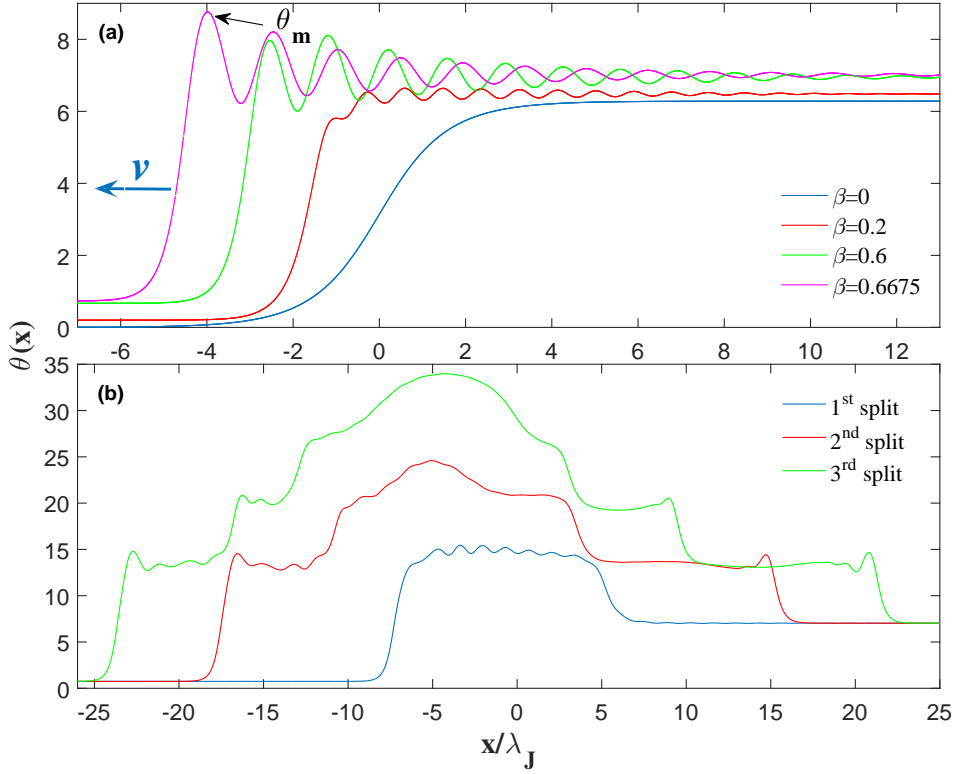}
\caption{\textbf{Steady-state vortex profiles and the initial stage of fragmentation instability.}  \textbf{a:} A sequence of phase profiles in a propagating vortex (shifted horizontally for clarity in the moving frames) calculated for a bulk junction by solving  equation (\ref{int}) for different values of $\beta$, $\eta=0.05$, and $\lambda_J/\lambda=10$. At $\beta_s=0.6676$ the peak amplitude of Cherenkov wave reaches $\theta_c=8.76$ and starts growing and evolving into an expanding vortex-antivortex pair. \textbf{b:} Snapshots of $\theta(x)$ at $\beta= \beta_s$ which show first three dissociations of the unstable vortex into vortex-antivortex pairs accompanied by Cherenkov radiation. Movies showing the initial stage of vortex instability and formation of the phase pile after multiple generations of vortex-antivortex pairs are available in Ref. \cite{suppl}. Notice that $\theta(\infty)-\theta(-\infty)=2\pi$ remains fixed by the phase difference in the initial vortex.}
\label{profiles}
\end{figure}

\break

\begin{figure}[b]
\centering\includegraphics[width=\textwidth]{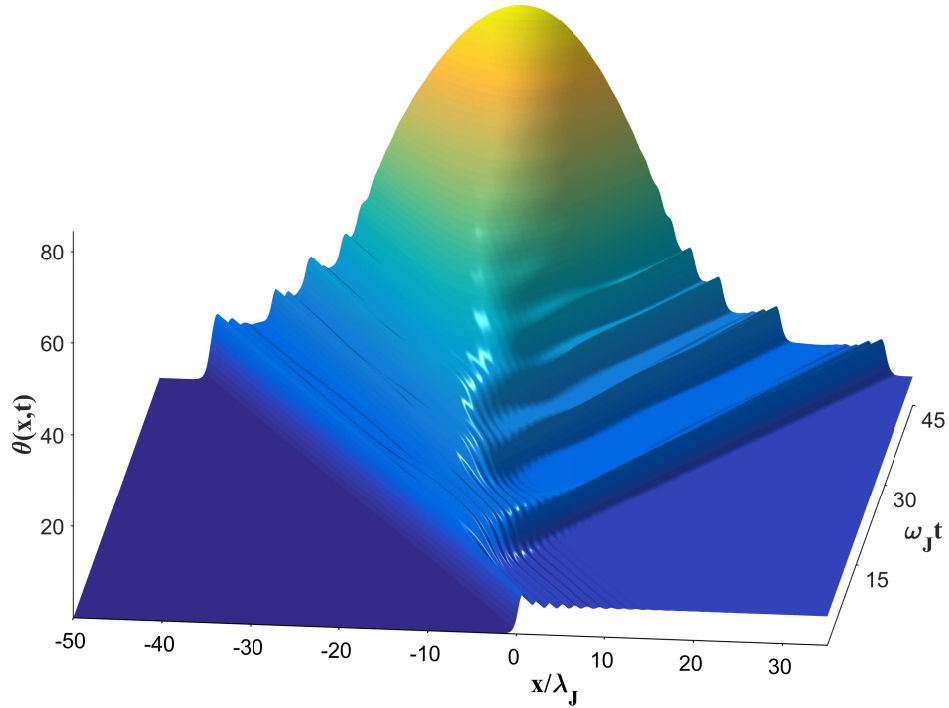}
\caption { \textbf{A 3D image of the evolution of phase pile triggered by an unstable vortex}. The dynamic phase distribution $\theta(x,t)$ was calculated from equation (\ref{int}) for a bulk junction at $\beta=0.6676$, $\lambda_J/\lambda=10$ and $\eta=0.05$. Here the maximum phase $\theta_m(t)$ increases approximately linear with time while the edge vortices move with constant velocities close to $c_s$. Individual vortices and antivortices clearly visible at the edges of the expanding phase pile overlap strongly toward its central part.}
\label{3d}
\end{figure}

\break

\begin{figure}[b]
\centering\includegraphics[width=\textwidth]{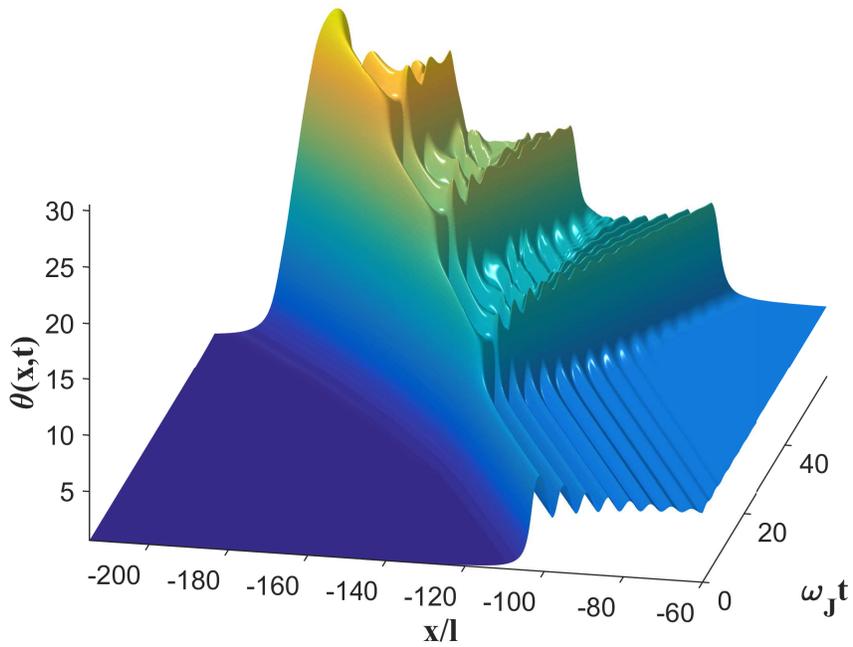}
\caption { \textbf{A 3D image of vortex instability and the initial stage of the phase pile formation in nonlocal regime}. The dynamic phase distribution was calculated at $\beta=\beta_s=0.63$ for a thin film edge junction with $\eta=0.1$ and $l=\Lambda/2$. Here $\theta(x,t)$ in the vortex at $\beta=\beta_s$ was computed by solving the full dynamic equation (\ref{int}) with the initial distribution equal to the stable single-vortex solution $\theta(x-vt)$ calculated at $\beta=\beta_s - 0.01$. As a result, the vortex then accelerates slightly and becomes unstable, triggering the growth of the phase pile. After multiple generations of vortex-antivortex pairs, vortices at the leading edges reach velocities of  $1.12l\omega_J$.}
\label{te3d}
\end{figure}

\break

\begin{figure}
\centering\includegraphics[width=\textwidth]{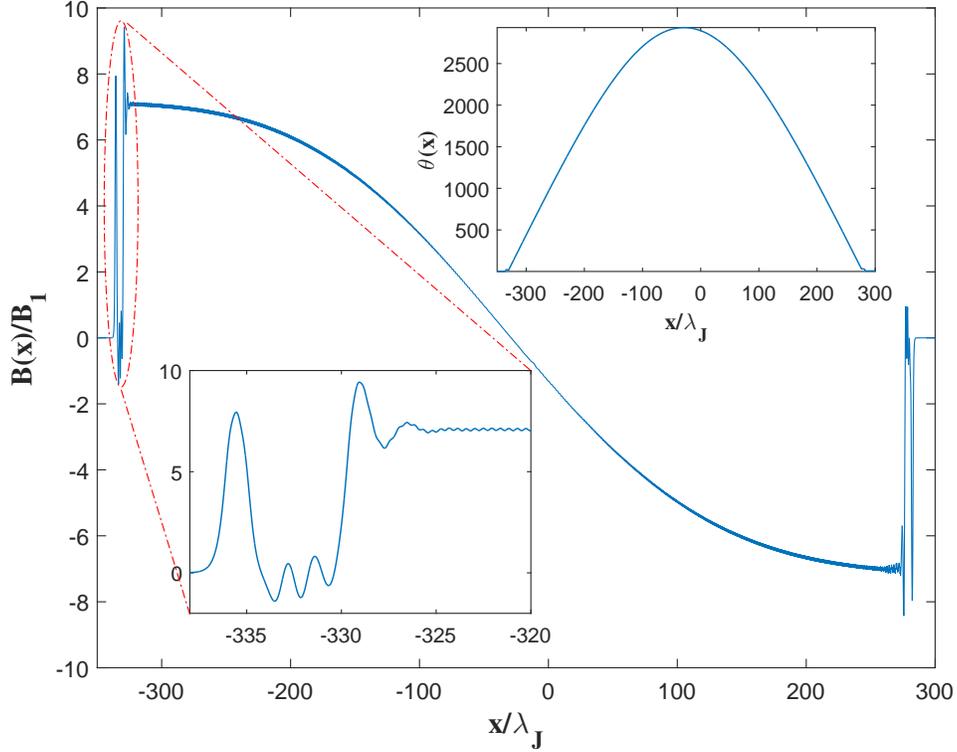}
\caption {\textbf{A snapshot of the normalized magnetic field $\mathbf{B(x,t)/B_1}$}. Here $B(x,t)$ was calculated from equation (\ref{int}) for a bulk junction at $\eta=0.05$, $\beta = 0.6676$, $\lambda_J/\lambda=10$ and $B_1=\phi_0/2\pi\lambda\lambda_J$. Inset shows the corresponding phase distribution, $\theta(x,t)$. One can clearly see a complex structure of the left leading edge comprised of a vortex overlapping with a vortex-antivortex pair. Away from the edges vortices overlap so strongly that the Cherenkov radiation gets suppressed almost to zero, and the smooth distribution of $B(x,t)$ in the growing resistive domain can be regarded as a giant multiquanta vortex-antivortex dipole.}
\label{field}
\end{figure}

\break

\begin{figure}
\centering\includegraphics[width=\textwidth]{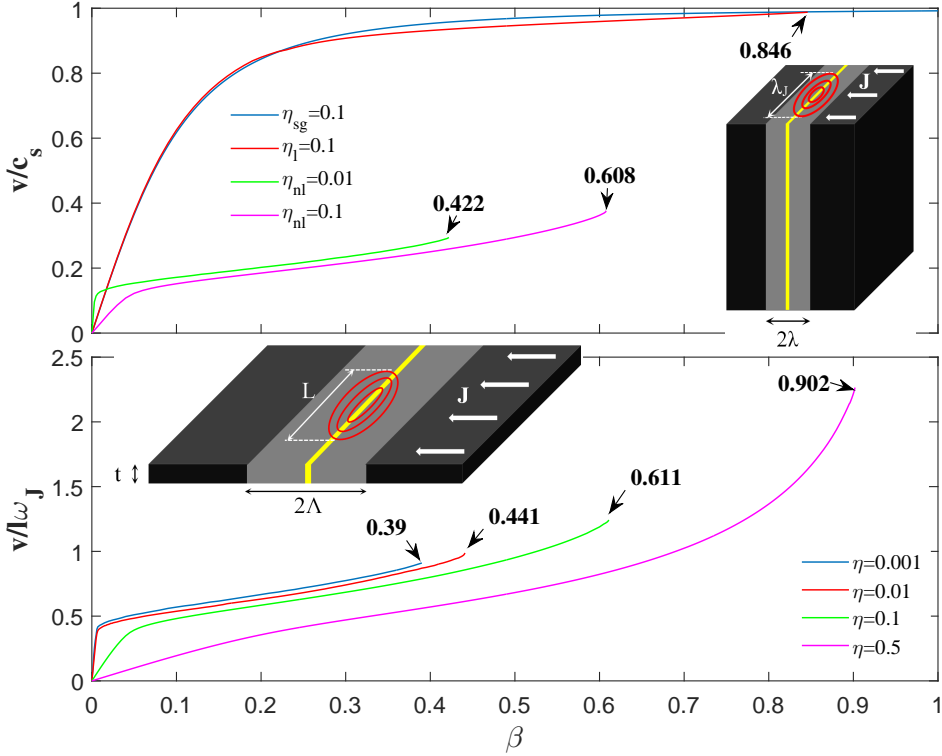}
\caption {\textbf{Velocities of a stable singe vortex $\mathbf{v(\beta)}$ as functions of current calculated for different junction geometries}. The instability occurs at the endpoints (shown by arrows) of the $v(\beta)$ curves. The upper panel shows $v(\beta)$ for a bulk junction calculated from equation (\ref{int}) at $\eta_l=0.1$ in the seemingly conventional weak-link limit, $\lambda_J/\lambda=10$ (for comparison, the blue curve shows $v(\beta)$ calculated from equation (\ref{sg}) at $\eta_{sg}=0.1$). The magenta and green curves show $v(\beta)$ calculated for a bulk nonlocal junction with $\lambda/\lambda_J=\sqrt{10}$ for values of $\eta_{nl}=0.1$ and $\eta_{nl}=0.01$, respectively, where the indices $sg$, $l$ and $nl$ correspond to the pure sine-Gordon, weakly nonlocal and strongly nonlocal limits, respectively. The lower panel shows results for a thin film edge junction in the extreme nonlocal limit described by equations (\ref{int}) and (\ref{log}). Notice that both the $v(\beta)$ curves and the critical values $\beta_s$ at $\eta=0.1$ and $\eta=0.01$ for the thin film junction are close to those for the bulk junction shown in the upper panel. This is because for a nonlocal bulk junction, $\theta''(u)$ in equation (\ref{int}) has a sharp peak of width $\sim (\lambda_J/\lambda)^2\lambda = 0.1\lambda$ so $G[(x-u)/\alpha]=\pi^{-1}K_0[|x-u|/\alpha]$ can be approximated by its expansion at small argument, $K_0(x/\alpha)\to \ln(2\alpha/|x|)-0.577$, which reduces to equation (\ref{log}). Here any constant factor under the $\log$ does not affect $\theta(x,t)$ since $\theta'(-\infty)=\theta'(\infty)=0$.  }
\label{iv}
\end{figure}

\break

\end{document}